\documentclass[gmd]{copernicus}
\usepackage{amsbsy}
\usepackage{amsmath}
\usepackage{amssymb}
\usepackage{verbatim}
\usepackage{xspace}
\usepackage{color}
\usepackage{enumitem}
\usepackage{listings}
\usepackage{setspace}
\usepackage[rightcaption]{sidecap}
\sidecaptionvpos{figure}{c}
\usepackage[]{hyperref}
\definecolor{DarkBlue}{rgb}{0.00,0.00,0.75}
\hypersetup{
  linkcolor   = DarkBlue,
  anchorcolor = DarkBlue,
  citecolor   = DarkBlue,
  filecolor   = DarkBlue,
  urlcolor    = DarkBlue,
  colorlinks  = true,
  pdftitle    = {Shingle 2.0: generalising self-consistent and automated domain discretisation for multiscale, unstructured mesh geophysical models},
  pdfauthor   = {Dr Adam S. Candy},
}

\usepackage{subfig}
\usepackage{cleveref}
\providecommand{\eqref}[1]{(\ref{#1})}
\providecommand{\shingle}{Shingle\xspace}
\providecommand{\libshingle}{LibShingle\xspace}

\providecommand{\opendap}{OPeNDAP\xspace}

\providecommand{\brml}{BRML\xspace}

\providecommand{\brep}{boundary representation\xspace}
\providecommand{\breps}{boundary representations\xspace}

\providecommand{\twod}{two-dimensional\xspace}
\providecommand{\threed}{three-dimensional\xspace}

\providecommand{\refdescriptionquotelink}{(\protect\hyperlink{description}{$\ast$})\xspace}
\providecommand{\refdescriptionquote}{($\ast$)\xspace}
\newenvironment{descriptionquote}{%
    \medskip\par\noindent%
    \hspace{0.005\columnwidth}%
    \begin{minipage}{0.91\columnwidth}%
    \itshape%
  }{%
    \end{minipage}%
    \hfill\refdescriptionquote%
    \medskip\par\noindent%
    \ignorespacesafterend%
  }

\crefname{equation}{}{}
\crefname{figure}{figure}{figures}

\usepackage{ntheorem}
\newtheorem*{constraint}{Constraints:}

\crefname{constraint}{constraints}{constraints}

\crefname{tenet}{tenet}{tenets}
\crefname{constraint}{constraint}{constraints}
\crefname{option}{option}{options}

\newlist{tenetenum}{enumerate}{1} 
\setlist[tenetenum]{label=\arabic*., ref=\arabic*}
\crefalias{tenetenumi}{tenet}

\newlist{constraintenum}{enumerate}{1} 
\setlist[constraintenum]{label=\emph{\arabic*.}, ref=\arabic*}
\crefalias{constraintenumi}{constraint}

\newlist{optionenum}{enumerate}{1} 
\setlist[optionenum]{label=\arabic*., ref=\arabic*}
\crefalias{optionenumi}{option}

\newcommand{\constraints}{\cref{constraint:brep,constraint:hmetric,constraint:id,constraint:surfbounds,constraint:vmetric}\xspace}

\usepackage{titlesec}
\titlespacing\section{0pt}{5pt plus 4pt minus 2pt}{2pt plus 2pt minus 2pt}
\titlespacing\subsection{0pt}{5pt plus 4pt minus 2pt}{2pt plus 2pt minus 2pt}
\titlespacing\subsubsection{0pt}{5pt plus 4pt minus 2pt}{2pt plus 2pt minus 2pt}
\usepackage[frozencache]{minted}

\newminted{python}{fontsize=\scriptsize,
               breaklines,
               xleftmargin=0.6em,
               linenos,
               numbersep=0.4em,
               frame=lines,
               framesep=2mm}
\newminted{xml}{fontsize=\scriptsize,
               breaklines,
               xleftmargin=0.0em,
               numbersep=0.2em,
               frame=lines,
               framesep=1mm}

\begin{document}
\setlength{\abovedisplayskip}{5pt plus 2pt minus 2pt}
\setlength{\belowdisplayskip}{5pt plus 2pt minus 2pt}
\nolinenumbers

\title{%
Shingle 2.0:
generalising self-consistent and automated
domain discretisation
for multi-scale
geophysical models${}^\dagger$
}

\runningtitle{
Shingle 2.0:
generalising self-consistent, automated
multi-scale geophysical domain discretisation
}
\author[1]{Adam~S.~Candy}
\author[1]{Julie~D.~Pietrzak}
\affil[1]{Environmental~Fluid~Mechanics~Section, Faculty~of~Civil~Engineering~and~Geosciences, Delft~University~of~Technology, The Netherlands}
\runningauthor{Candy, A.S.}
\correspondence{A.S.~Candy~(a.s.candy@tudelft.nl)
\\${}^\dagger$Library code, verification tests and examples available in the repository at \url{https://github.com/shingleproject/Shingle}.
Further details of the project presented at \url{https://www.shingleproject.org}.
}
\received{}
\revised{}
\accepted{}
\published{}
\firstpage{1}
\maketitle

\begin{abstract}
The approaches taken to describe and develop spatial discretisations of the domains required for geophysical simulation models are commonly ad hoc,
model or application specific and
under-documented.
This is particularly acute for simulation models that are flexible in their use of multi-scale, anisotropic, fully unstructured meshes where a relatively large number of heterogeneous parameters are required to constrain their full description.
As a consequence, it can be difficult to reproduce simulations, ensure a provenance in model data handling and initialisation, and a challenge to conduct model intercomparisons rigorously.

This paper
takes a novel approach to spatial discretisation, considering it much like a numerical simulation model problem of its own.
It introduces a generalised, extensible, self-documenting approach to carefully
describe, and necessarily fully, the constraints over the
heterogeneous parameter space that determine how a domain is spatially discretised.
This additionally provides a method to accurately record these constraints,
using high-level natural language based abstractions,
that enables full accounts of provenance, sharing and distribution.
Together with this description, a generalised consistent approach to unstructured mesh generation for geophysical models is developed,
that is automated, robust and repeatable, quick-to-draft, rigorously verified and consistent to the source data throughout.
This interprets the description above to execute a self-consistent spatial discretisation process,
which is automatically validated to expected discrete characteristics and metrics.
\end{abstract}

\introduction
\label{sec:introduction}
\begin{figure*}[!h]
\begin{center}
\includegraphics[width=\textwidth]{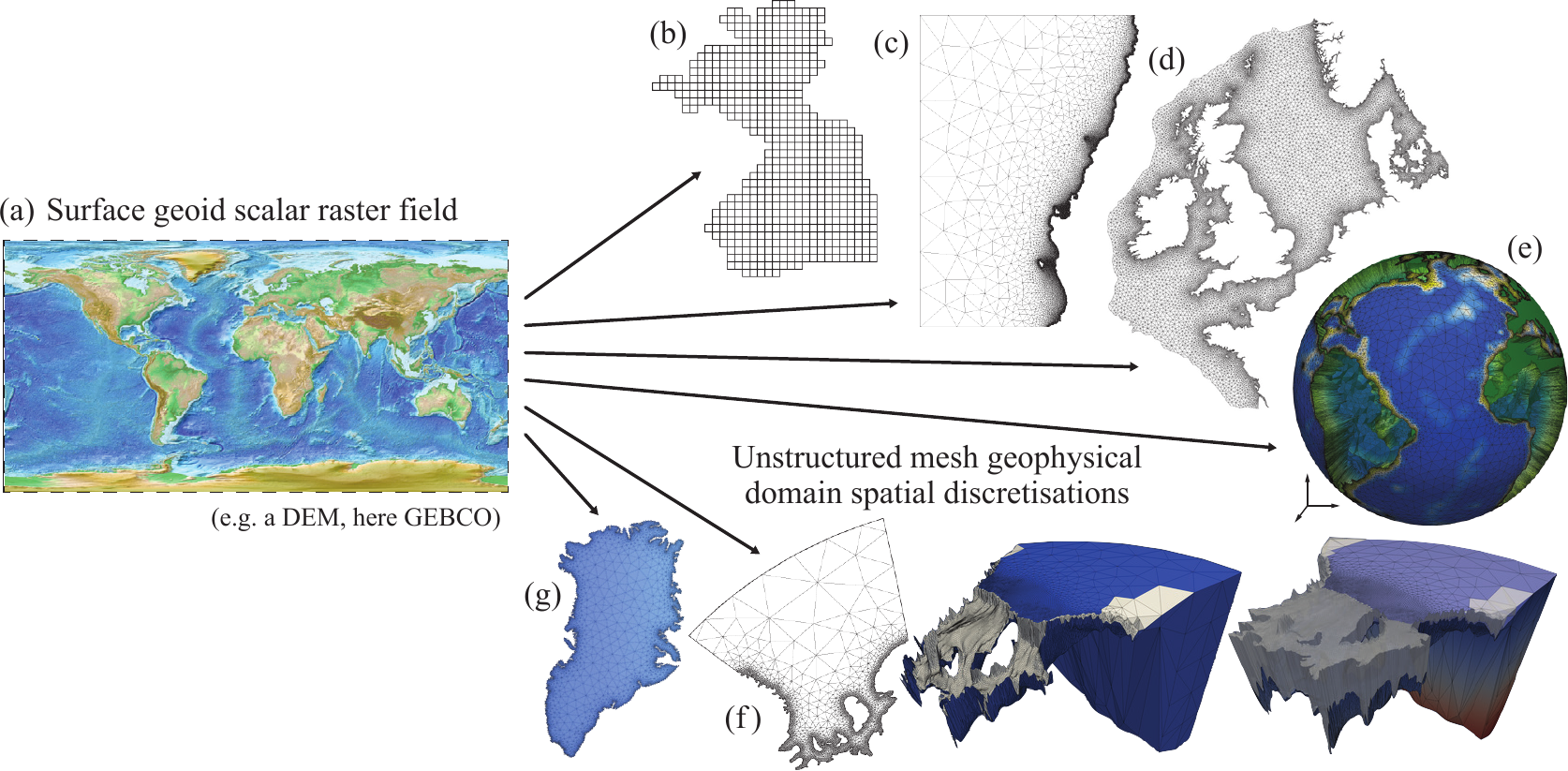}
\end{center}
\vspace{-1.8ex}
\caption{%
The challenge:
to generate a self-consistent domain discretisation approach for geophysical domains
that is generalised such that it can be applied to a wide range of applications,
with new domains efficiently prototyped and iterated on,
and is fully described such that
the process can be automated,
is reproducible
and easily shared.
(a) shows a typical source Digital Elevation Map (DEM) dataset
(that naturally lend themselves to structured grid generation)
used to produce a regular grid of the Atlantic Ocean (e.g. under a format-native land mask) in (b),
and a selection of
unstructured mesh spatial discretisations:
(c) Bounded by part of the Chilean coastline and a meridian.
(d) North Sea.
(e) Global oceans.
(f) Grounding line of the Filchner-Ronne ice shelf ocean cavity up to the 65\degree S parallel, with
surface geoid mesh $\mathcal{T}_h$,
full mesh $\mathcal{T}$ with ice-ocean melt interface highlighted,
and accompanied by ice sheet full discretisation.
(g) Greenland ice sheet.
}
\label{fig:challenge}
\end{figure*}
\begin{figure*}[!h]
\begin{center}
\includegraphics[width=\textwidth]{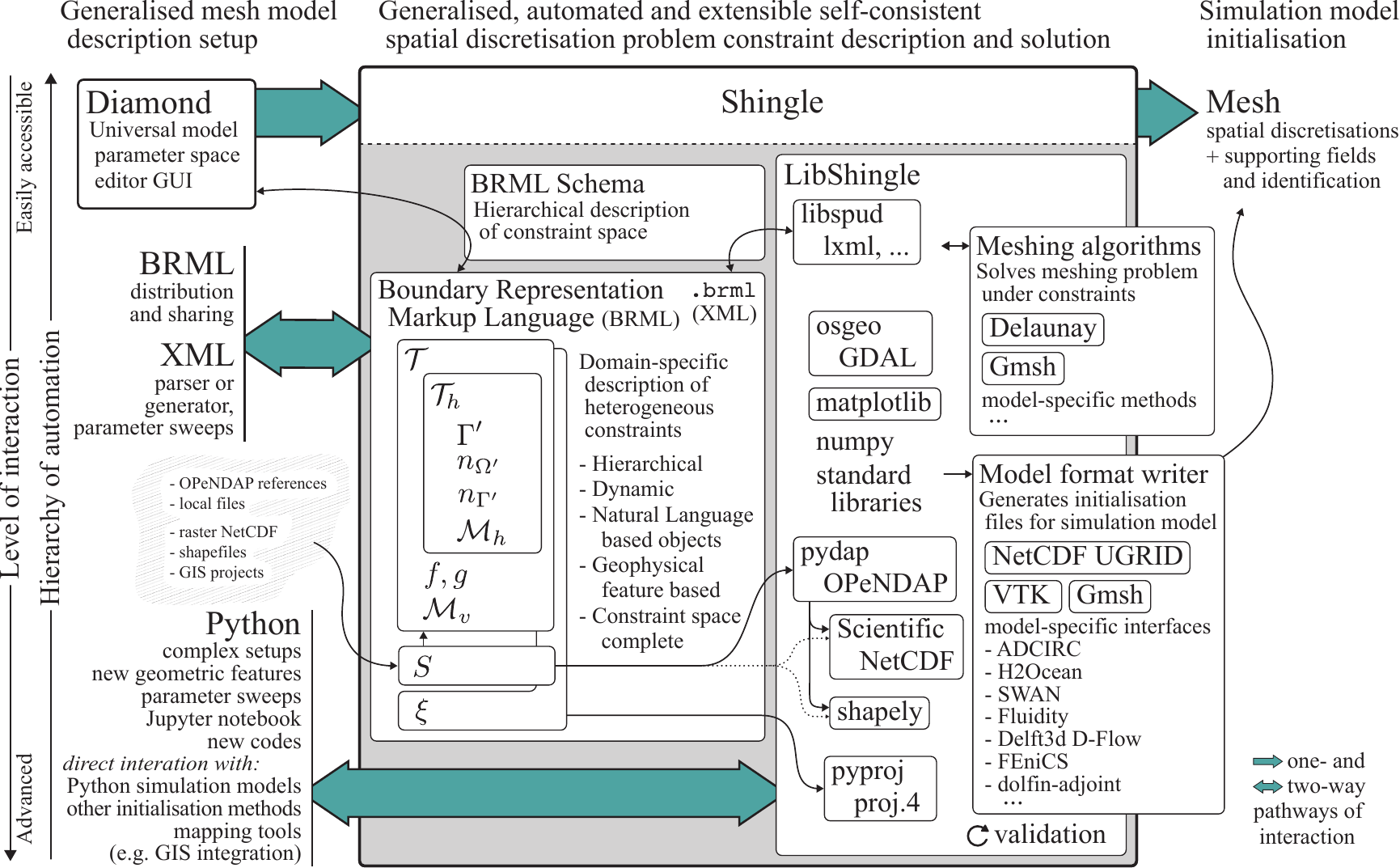}
\end{center}
\vspace{-1.8ex}
\caption{%
A schematic illustrating the generalised approach to flexible unstructured mesh specification and generation for geophysical models.
The hierarchy of automation (\cref{tenet:automated}) is highlighted, from
a relatively simple high-level interaction:
{Diamond GUI $\leftrightarrow$ Shingle $\rightarrow$ Mesh},
to complex low-level development communicating with the \libshingle library.
Nomenclature defined in \cref{sec:meshgeneration}.
}
\label{fig:process}
\end{figure*}

Numerical simulation models have become a vital tool
for scientists studying geophysical processes.
Mature operational models inform continuously updated short-term public weather forecasts,
whilst studies of mantle dynamics and ice sheet evolution improve understanding of physical systems in relatively inaccessible locations, where data is sparse.

Use of unstructured mesh spatial discretisations%
\footnote{For the purposes of the discussion here, \emph{spatial discretisation} specifically refers to the division of a continuous spatial domain into discrete  parts -- a discrete tessellation or honeycomb -- a generalised notion of triangulation.}
is growing in the fields of modelling geophysical systems,
where it is possible to conform accurately to complex, fractal-like surfaces and
vary spatial resolution to optimally capture the physical process, or multi-scale range of processes under study.
The past few years have seen
a global unstructured ocean model
\citep[FESOM,][]{sidorenko14}
join structured studies in
internationally coordinated climate studies,
such the
Coupled Model Intercomparison Project \citep[CMIP,][]{meehl07,taylor12}
and
the
Coordinated Ocean-ice Reference Experiments
\citep[CORE][and accompanying studies in the \emph{Ocean Modelling} special issue]{coreii}.
More are in active development \citep[e.g.][]{ringler13}
and the number of unstructured models joining these efforts
--
that directly contribute to
reports compiled by
the Intergovernmental Panel on Climate Change (IPCC)
--
likely to grow.
Similarly, on smaller scales,
the geometric flexibility of unstructured discretisations are being
applied
to reduce the need for nesting models,
and in accurately applying forcings or coupling physics
\citep[e.g.][]{kimura13}
on complex and possibly dynamic, deformable physical interfaces.
At the cusp where these efforts meet, prospects for
introducing successively greater complexity
in the representation of coastal seas in global ocean models are reviewed in
\cite{holt17}.

The challenge (see \cref{fig:challenge})
of constraining and fully describing an
arbitrarily unstructured
spatial discretisation
bounded by complex, fractal-like bounds that typically characterise geophysical domains,
with inhomogeneous and potentially anisotropic spatial resolution,
is a significant one.
Defining the domain geoid bounds is no longer a simple case of applying a land mask to similarly regular gridded data.
The generalised constraints are now a heterogeneous set of functions \citep{candybrep},
and as a consequence are more difficult to describe.
In general, domain discretisations are often under-described leaving it difficult to repeat simulations exactly,
which particularly for the unstructured case, can have a strong influence on model output.
Not only is the description and generation process a significant challenge, but achieving this in a way
that maintains a record of provenance such that simulations as a whole are reproducible,
that scales and is efficient, and consistent to source data
-- attributes required and expected in scientific modelling studies --
make this a much more difficult problem (summarised in \cref{fig:tenets}).
Existing, standard structured-mesh tools cannot be used.

Grid generation for geophysical models in real domains is not only becoming a significantly more complex and challenging problem
to constrain and describe,
but additionally in the computational processing required.
As models include a greater range of spatial scales, more computational effort is required to optimise the discretisation before a simulation proceeds
(e.g. the actively developed MPAS models, \cite{ringler13}, strongly optimise their hexagonal prism based mesh discretisation).
An increasing number of geometric degrees of freedom demand the meshing process is broken up over multiple parallel threads
\citep[as demonstrated in][]{candybrep}, just as simulation models have evolved to run in parallel.

These challenges are identified in \cite{candybrep} by the \emph{nine tenets of geophysical mesh generation}, summarised in \cref{fig:tenets}.
This work takes the view that significant progress can be made towards these by approaching the mesh generation problem in the same way as a numerical simulation model.

Simulation domains in geophysical models are typically defined with reference to geographical features.
A tsunami simulation geoid surface domain is, for example, usually described by a length of coastline between two points
(commonly marked by longitude or latitude references) extended out to an orthodrome.
In the case of 2010 Chile earthquake
centred about
35.9\degree S 72.7\degree W
(see \cref{fig:chile}),
the domain is concisely described:

\begin{descriptionquote}
\hypertarget{description}{}
``
\scalebox{0.90}[1.0]{... \!bounded by the 0m depth coastline from 32\degree \!S to 40\degree \!S,}
\phantom{``... \!} extended along parallels to the 77\degree \!W meridian, \\
\phantom{``... \!} in a latitude-longitude WGS84 projection \!...''
\label{description}
\end{descriptionquote}

As part of the generalisation of domain description, this new approach interacts directly with these natural language based
geographic
references,
structured by a formal grammar,
to provide a general, model-independent and accurate description of spatial discretisation for geophysical model domains.
This forms part of the \cite{shingle} computational research software library, that accompanies this work, providing a novel approach to describing and generating highly multi-scale boundary-conforming domain discretisations, for seamless concurrent simulation.

The objective of this paper is to provide:
\begin{enumerate}
\vspace{-1.0ex}
\setlength{\itemsep}{1pt}\setlength{\parskip}{0pt}\setlength{\parsep}{0pt}
\item A user-friendly, accessible and extensible framework for model-independent geophysical domain mesh generation.
\item An intuitive, hierarchical formal grammar to fully describe and share the full heterogeneous set of constraints for the spatial discretisation of geophysical model domains.
\item Natural language basis for describing geophysical domain features.
\item Self-consistent, scalable, automated and efficient mesh prototyping.
\item Platform for iterative development that is repeatable, reproducible with a provenance history of generation.
\end{enumerate}
\vspace{-1.0ex}
With significant progress made through the novel approach of considering the problem much like that of a numerical simulation model problem.

The previous work \cite{candybrep}
developed a consistent approach to domain discretisation,
with a focus on uniform processing and data sources, which further enabled the discretisation of domains not possible with standard approaches.
Additionally, it identified the complete set of heterogeneous constraints required to fully describe a mesh generation problem for the discretisation of geophysical domains.
This work now extends and generalises this consistent approach introducing a natural language based formal grammar for a modeller to describe and share the constraints.
Under the formal grammar the description is ensured necessarily complete,
such that the problem is fully constrained and is therefore reproducible.
This employs the
novel hierarchical problem descriptor
framework
Spud \citep{ham09} which has been specifically designed to manage large and diverse option trees for numerical models.
The formal self-describing data file is a universal, shareable description of the full constraints, written in a standard data format, presented in context through a natural hierarchical structure, readable by established open source libraries.

The pathways of interaction with the library have grown
(outlined in \cref{fig:process}),
such that it is accessible to a wide range of users.
Its modular library framework, with for example,
geospatial operations, homeomorphic projections, meshing algorithms and model format writers
are the focus of distinct modular parts,
and the use of standard external libraries where possible,
allows development to remain in small sections of the code base,
such that develops can stay within their specialisms.
Additionally, the dictionary approach to managing option parameters taken by Spud means new features can be added and exposed through interfaces, such as the Diamond Graphical User Interface (GUI), without the need to pass new arguments through code functions,
and similarly require small changes and only in low-level code.

Output writers in the library prepare the solution discretisation for use in simulation codes,
in cases where the output Python objects cannot be used directly,
encouraging the use of standard formats
and also supporting existing proprietary model-specific formats.
These additionally support supplementing the spatial discretisation
(which itself includes a vector field describing mesh node coordinate locations)
with additional interpolated fields for simulation model initialisation and forcing
(\cref{fig:process}).

Through both the objects in the problem description file
(\cref{fig:block,fig:gui,fig:brml})
and those in the Python library \libshingle
(\cref{fig:python}),
\shingle provides a
language to combine geographic components to build up boundary representation, mesh spatial variation and identification
-- a high-level abstraction to the complex constraint description problem --
which is then processed by the library in deterministic (or as close to as possible) process to accurately construct the specified mesh
in a repeatable way.

The validation tests of \cite{candybrep} have been significantly widened from the limited boundary representation
tests
to include expected
discrete properties and metrics of the high fidelity description and resulting domain discretisation.
These expected characteristics are prescribed as part of the self-describing problem file, such that other users can check the output is as intended.
This self-contained description and validation is then straight-forwardly processed by the library verification engine, making it easy to add new tests.

Through this approach,
geophysical domain discretisation can be the relatively simple steps (top of \cref{fig:process}) of using the Diamond GUI to choose a dataset and specify bounds using natural language objects,
which is then run through the Shingle executable to produce a mesh.
This is accessible and straightforward to new users.
More so with the suite of test cases that provide examples and easily ensure verification through a built-in test engine.

More advanced use can be built up in stages through the GUI,
with validation checks on expected mesh properties easily added
to ensure reliable reproduction
throughout the iterative mesh prototyping process.
Beyond this the XML based description is easily interrogated and modified with standard tools.
Lower-lever still, the natural language based objects and discretisation constraints can be accessed directly through its Python library interface.
This has grown since its first iteration reported in \cite{candybrep},
where it was used to develop complex discretisations dependent on the mean position of
Antarctic Circumpolar Current (ACC) and domains to complex grounding line positions under the floating ice shelves of Antarctica.
Python plugins for QGIS \citep{qgismanual} were developed using parts of the \shingle library code to demonstrate integration with Geographic Information Systems (GIS) in \cite{candygis}.

With mesh generation becoming a complex problem to describe and a computationally challenging process,
that we argue is best handled in an approach that mirrors the development of a numerical simulation model,
support and interaction with other frameworks such as GIS is best maintained
with a standalone library and a formal problem description specifically designed to constrain the general geophysical domain discretisation problem.

The paper is structured such that the following
\cref{sec:meshgeneration}
sets out the challenge, reviewing the set of heterogeneous \constraints required to fully describe a domain discretisation problem, and key considerations in \cref{fig:tenets}.
The natural language based \brml problem description is introduced in
\cref{sec:brml}, with a consideration of source data in \cref{sec:data}.
The \libshingle library central to the generalised approach (illustrated in \cref{fig:process}) is detailed in
\cref{sec:framework}
with ways to interacting with the framework presented in
\cref{sec:interaction}.
Examples and validation is covered in
\cref{sec:verification},
with conclusions made in
\cref{sec:conclusion}.

\section{Generalised unstructured spatial discretisation for geophysical models}
\label{sec:meshgeneration}
\subsection{Constraints for mesh generation in geophysical domains}
\label{sec:constraint}
The contrast in dominant dynamical processes that characterise geophysical systems,
split in orthogonal directions parallel and perpendicular to
the local gravitational acceleration $\boldsymbol{g}$,
leads to a spatial decoupling that
restricts the parameter space of general spatial domains
$\Omega \in \mathbb{R}^3$.
Meshes of geophysical domains can be built differently in these distinct directions in order to well-support the associated dynamics,
with mesh characteristics on the geoid plane
considered independently of those in the perpendicular direction of $\boldsymbol{g}$.
A formal description of the heterogeneous set of constraint functions, homeomorphic mappings and topological spaces,
required to fully describe geophysical model domain spatial discretisations,
is developed and detailed in \cite{candybrep},
of which a summary of the key outcome follows.
\begin{constraint}
\label{constraints}
The spatial domain discretisation for a computational geophysics simulation in a domain $\Omega \subset \mathbb{R}^3$,
requires the constraint of
\begin{constraintenum}[leftmargin=1.2em]
\renewcommand{\emph}[1]{\textbf{#1}}
\item \label{constraint:brep} \emph{Geoid boundary representation}
$\Gamma_g$,
of the geoid surface $\Omega_g \subset \mathbb{R}^3$,
inclusive of the maximal extent of $\Omega$ perpendicular to $\boldsymbol{g}$.
Under a homeomorphic projection
$\xi$,
this is considered as
the chart
$\Omega' \subset \mathbb{R}^2$,
such that the boundary $\Gamma'$ is described by
\begin{flalign}
\quad
\Gamma'\!:
t \in \mathbb{R}
\mapsto
\zeta(t) \in \mathbb{R}^2,
&&
\label{brep}
\end{flalign}
an orientated vector path of the encompassing surface geoid bound defined in \twod parameter space. 
\item \label{constraint:hmetric} \emph{Geoid element edge-length resolution metric} for dynamics aligned locally to a geoid, described by the functional
\begin{flalign}
\quad
\mathcal{M}_h\!:
\boldsymbol{x} \in
\Omega'
\mapsto
\mathcal{M}_h (\boldsymbol{x}) \!\in
\mathbb{R}^2\!\times\!\mathbb{R}^2\!.
&&
\label{hmetric}
\end{flalign}
\item \label{constraint:id} \emph{Boundary and region identification}, prescribed by
\begin{flalign}
\quad
n_{\Gamma'}\!:
t \in \mathbb{R}
\mapsto
n_{\Gamma'} (t) \in \mathbb{Z},
\textrm{ and}
&&
\label{idbound}
\end{flalign}
\vspace{-2em}
\begin{flalign}
\quad
n_{\Omega'}\!:
\boldsymbol{x} \in \Omega'
\mapsto
n_{\Omega'} (\boldsymbol{x}) \in \mathbb{Z},
\textrm{ respectively.}
&&
\label{idregion}
\end{flalign}
\item \label{constraint:surfbounds} \emph{Surface bounds}, height maps defined on the surface geoid domain, described by the functions
\begin{flalign}
\quad
f,g\!:
\boldsymbol{x}
\mapsto
\mathbb{R}
\quad
\forall
\boldsymbol{x} \in \Omega'.
&&
\label{surfbounds}
\end{flalign}
\item \label{constraint:vmetric} \emph{Vertical element edge-length resolution metric} for dynamics in the direction of gravitational acceleration
(e.g. buoyancy driven),
described by the functional
\begin{flalign}
\quad
\mathcal{M}_v\!:
\boldsymbol{x} \in
\Omega
\mapsto
\mathcal{M}_v (\boldsymbol{x}) \in
\mathbb{R}.
&&
\label{vmetric}
\end{flalign}
\end{constraintenum}
\end{constraint}

\subsection{Decoupled mesh development}
\label{sec:decoupled}
The spatial decoupling permits discretisation in two stages
corresponding to directions parallel and perpendicular to the local gravitational acceleration (refer to \cref{fig:block}).
Firstly, the `horizontal' geoid surface domain discretisation problem is solved
under \cref{constraint:brep,constraint:hmetric,constraint:id}
using
the surface geoid boundary representation $\Gamma'$~\eqref{brep},
geoid element edge-length metric $\mathcal{M}_h$~\eqref{hmetric},
with
boundary and region identifications,
$n_{\Gamma'}$~\eqref{idbound}
and
$n_{\Omega'}$~\eqref{idregion}
respectively,
such that
\begin{equation}
h\!: \{\Gamma', \mathcal{M}_h, n_{\Gamma'}, n_{\Omega'}\} \mapsto \mathcal{T}_h,
\label{h}
\end{equation}
a tessellation of $\Omega' \subset \mathbb{R}^2$, with identification elements.

Secondly, if needed, this is followed by discretisation in a direction aligned with gravitational acceleration.
The
\cref{constraint:surfbounds,constraint:vmetric},
describing the
surface bounds $f$ and $g$~\eqref{surfbounds}
and
vertical edge-length metric $\mathcal{M}_v$~\eqref{vmetric},
together with
the surface geoid discretisation $\mathcal{T}_h$ \cref{h},
forms a discretisation problem that is solved through the process
\begin{equation}
v\!: \{\mathcal{T}_h, f, g, \mathcal{M}_v\} \mapsto \mathcal{T},
\label{v}
\end{equation}
to give the full domain discretisation of $\Omega \subset \mathbb{R}^3$, with identification elements.

\subsection{The nine tenets of geophysical mesh generation}
\label{sec:tenets}
\begin{table}[!h]
\centering
\begin{minipage}{\columnwidth}
{
\renewcommand{\emph}[1]{\textbf{\textsf{#1}}}
\noindent \hrulefill
\begin{tenetenum}[
labelindent=*,
style=multiline,
leftmargin=*,
]
\setlength\itemsep{0em}
\vspace{-0.5em}
\item \label{tenet:brep} 
      Accurate description and \emph{representation of arbitrary and complex boundaries}
      such that they are contour-following to a degree prescribed by the metric size field,
      with aligned faces so forcing data is consistently applied ($\Gamma'$, $f$, $g$).
      
\item \label{tenet:metric} 
      \emph{Spatial mesh resolution} to minimise error; with efficient aggregation of contributing factors,
      ease of prototyping and experimentation of metric functions and contributing fields,
      over the entire extent of the bounded domain ($\mathcal{M}_h$, $\mathcal{M}_v$).
\item \label{tenet:region} 
      Accurate geometric \emph{specification of regions} and \emph{boundary features};
      to provide for appropriate interfacing of regions of differing physics,
      model coupling and parameterisation application ($n_{\Omega'}$, $n_{\Gamma'}$).
\item \label{tenet:consistent} 
      \emph{Self-consistent}, such that all contributing source data undergoes the same pre-processing,
      ensuring self-consistency is inherited.
      
\item \label{tenet:efficient} 
      \emph{Efficient drafting and prototyping} tools,
      
      such that user time can be focused on high-level development of the physics and
      initialisation of the modelled system.

\item \label{tenet:scales} 
      \emph{Scalability}, with operation on both small and large datasets,
      facilitating the easy manipulation and process integration, independent of data size.

\item \label{tenet:automated} 
      \emph{Hierarchy of automation}, such that individual automated elements of the workflow
      can be brought down to a lower-level for finer-scale adjustments.
\item \label{tenet:provenance} 
      \emph{Provenance} to ensure the full workflow from initialisation to simulation and
      verification diagnostics are reproducible.
\item \label{tenet:standard} 
      \emph{Standardisation of interaction} to enable interoperability between both tools and scientists.
\vspace{-1em}
\end{tenetenum}
\noindent \hrulefill
\vspace{-1ex}
}
\end{minipage}
\caption{
The nine tenets of geophysical mesh generation.
Solutions to the spatial discretisation of geophysical model domains need to address these nine attributes
\cite[from][]{candybrep}.
}
\label{fig:tenets}
\end{table}
Accompanying the constraints, \cite{candybrep} identifies the nine attributes listed in \cref{fig:tenets} as key to geophysical mesh generation processes.

\section{Boundary Representation Markup Language}
\label{sec:brml}
\subsection{Unstructured domain discretisation a model problem}
The functional forms \cref{brep,hmetric,idbound,idregion,surfbounds,vmetric}
of the unstructured meshing problem
require a range of types of data,
from more standard \twod raster maps, to tensors and orientated vector paths.
It is a challenge to manage this heterogeneous collection of parameters (\cref{tenet:efficient,tenet:provenance}),
such that they are handled consistently (\cref{tenet:consistent})
and for the level of complexity that can be encountered (\cref{tenet:scales,tenet:automated}).
This is in contrast with the structured mesh case, which requires relatively simple data of the same format as its
inputs:
a \twod Digital Elevation Map (DEM) raster dataset supplying a \twod raster mask, for example.

\begin{figure}[!h]
\vspace*{2mm}
\begin{center}
\includegraphics[width=\columnwidth]{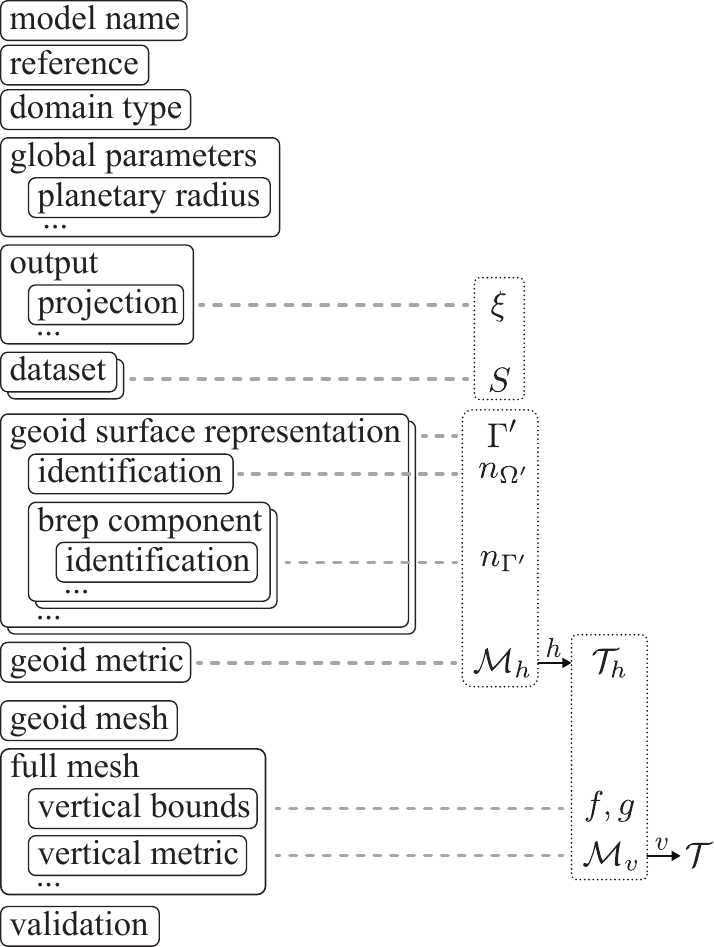}
\end{center}
\vspace{-1.8ex}
\caption{%
Overview layout of
geophysical domain mesh
constraint description
highlighting extensible dynamic components
and correspondence to
source data $S$, projection $\xi$ and
\cref{constraint:brep,constraint:hmetric,constraint:id,constraint:surfbounds,constraint:vmetric}.
}
\label{fig:block}
\end{figure}

\begin{figure}[!h]
\begin{center}
\includegraphics[width=\columnwidth]{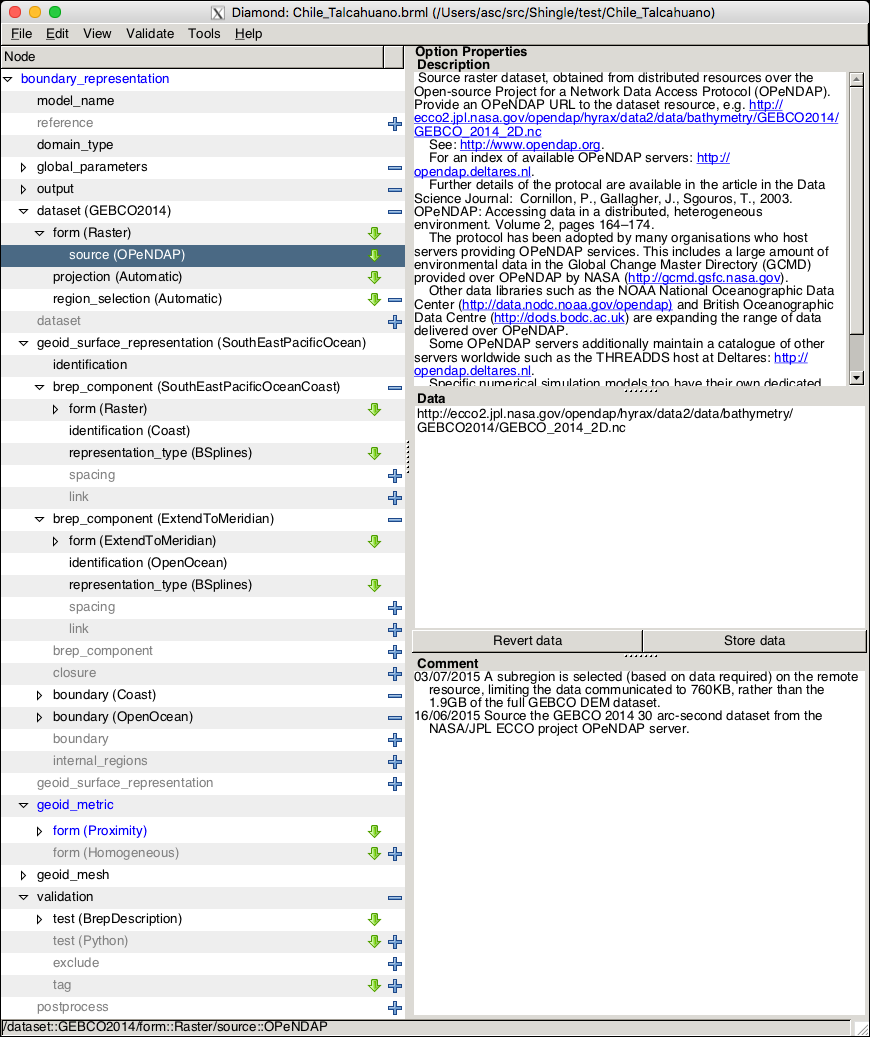}
\end{center}
\vspace{-1.8ex}
\caption{%
An example view of the Graphical User Interface Diamond
inspecting the
hierarchical tree of option parameters that fully constrain the
geophysical domain mesh problem.
Each node is shown in context on the left,
with their
option properties presented on the right,
including
raw data
and the possibility to
note comments.
This is guided by the \brml schema
developed and supplied with \shingle,
which additionally provides the fuller
self-describing
option descriptions
shown in the top right.
Options down the tree highlighted in blue are
mandatory
and guide the user to defining a complete set of constraints.
}
\label{fig:gui}
\end{figure}
\begin{figure}[]
\centering
\begin{xmlcode}
<?xml version='1.0' encoding='utf-8'?>
<boundary_representation>
  <model_name>
    <string_value lines="1">Chile_Talcahuano</string_value>
  </model_name>
  <global_parameters/>
  <output>
    <projection>
      <string_value>LatLongWGS84</string_value>
    </projection>
  </output>
  <dataset name="GEBCO2014">
    <form name="Raster">
      <source name="OPeNDAP" file_name="http://ecco2.jpl.nasa.gov/ opendap/hyrax/data2/data/bathymetry/GEBCO2014/ GEBCO_2014_2D.nc"/>
    </form>
    <projection name="Native"/>
    <region_selection name="Automatic"/>
  </dataset>
  <geoid_surface_representation name="SouthEastPacificOcean">
    <identification>
      <integer_value rank="0">9</integer_value>
    </identification>
    <brep_component name="SouthEastPacificOceanCoast">
      <form name="Raster">
        <source name="GEBCO2014"/>
        <region>
          <longitude>
            <minimum>-77.0</minimum>
            <maximum>-71.0</maximum>
          </longitude>
          <latitude>
            <minimum>-40.0</minimum>
            <maximum>-32.0</maximum>
          </latitude>
        </region>
        <contourtype name="coastline0m"/>
        <comment>Simple single bounding box centred about the epicentre 35.909°S 72.733°W.</comment>
      </form>
      <identification name="Coast"/>
      <representation_type name="BSplines"/>
    </brep_component>
    <brep_component name="OpenMeridian">
      <form name="ExtendToMeridian">
        <longitude>
          <real_value rank="0">-77.0</real_value>
        </longitude>
      </form>
      <identification name="OpenOcean"/>
      <representation_type name="BSplines"/>
    </brep_component>
    <boundary name="Coast">
      <identification_number>
        <integer_value rank="0">3</integer_value>
      </identification_number>
    </boundary>
    <boundary name="OpenOcean">
      <identification_number>
        <integer_value rank="0">4</integer_value>
      </identification_number>
    </boundary>
  </geoid_surface_representation>
  <geoid_metric>
    ...
  </geoid_metric>
  <validation>
    <test name="BrepDescription" file_name="data/Chile_Talcahuano.geo.bz2">
      <compressed/></test>
    <test name="NodeNumber"> ... </test>
  </validation>
</boundary_representation>
\end{xmlcode}
\vspace{-3ex}
\caption{
Example domain discretisation description, in a self-describing
\brml
description file
(with a few parts marked ... skipped).
This is a human-readable translation of the simple description
\refdescriptionquotelink
under the formal grammar of the schema that defines the geophysical domain discretisation constraint space.
This file is examined by the GUI in \cref{fig:gui} and,
on straight-forward and automated processing by \shingle, produces the
simulation-ready spatial discretisation of \cref{fig:chile}.
}
\label{fig:brml}
\end{figure}

Mesh specification in the unstructured case,
with flexibility to include conforming boundaries,
is much more like the initialisation of a numerical simulation model.
This typically contains a heterogeneous set of functions:
those defined over
$\mathbb{R}^3$ initialising or forcing full fields, together with boundary conditions defined on surfaces in $\mathbb{R}^2$ and
potentially line and point sources, or full field functions of reduced rank such as the gravitational acceleration parameter,
or value of a bulk eddy viscosity, for example.
Mesh descriptions and constraints are only going to become more complex as simulation models include a larger range of spatial scales and physical processes.
Moreover, like a simulation model, unstructured mesh generation includes calculations that can be computationally demanding.
The generation of conforming boundary representations is no longer a simple binary operation identifying which elements lie in the simulation domain through mask fields.
Similarly, the construction of domain discretisations with variable element sizes contains many more unknowns
in the unstructured case than the corresponding
local cell-division approaches typically used to increase spatial resolution in the structured case.

In light of this,
\shingle takes the approach that domain discretisation specification and generation is best considered as a model problem.
Formalised, the output mesh is the solution of a discretisation problem under a heterogeneous parameter space of constraints.

\subsection{Spud constraint space management}
\label{sec:spud}
Much like numerical model input parameter specification, mesh generation is often overlooked, and a secondary consideration to the dynamical core of a numerical model.
Typically inputs are ad hoc, model-specific, plain text files containing name lists that are expanded as a model develops.
For only but simple cases, this leads to model interfaces (and their associated pre- and post-processing tools) that are difficult to maintain
and simulation setups that are not easily shared and understood.

This problem of model input parameter specification is considered in \cite{ham09},
together with the proposed solution Spud.
This provides a generalised, model-independent method of describing all constraints to a model problem,
that is dynamic, easily extensible with a hierarchical context for parameters.
Formal grammars guide user input, minimise errors and formalise parameter specification.

\subsection{Constraint space description}
The options
available to describe a mesh discretisation
are typically defined by model interfaces.
These tend to be ad hoc and unportable, tied directly to numerical simulation codes.
Initialisation tools then require their own implementation to interpret and write model options, which is prone to error and potential inconsistencies.

Existing file formats have been used,
and their syntax overloaded,
to describe geophysical spatial discretisations.
Ice sheet domains are built up using a Constructive Solid Geometry (CSG) approach within
the COMSOL~\citep{comsol,li09} multi-physics modelling environment in \cite{humbert09}.
The GeoCUBIT~\citep{geocubit} branch of CUBIT developed for seismic inversion domains,
and a plugin for Gmsh~\citep{gmsh} to enable the creation of domains bounded by paths from
the Global Self-consistent, Hierarchical, High-resolution Geography \cite[GSHHG,][]{gshhs}.
Extensions to GIS \citep[e.g.][]{candygis} enable a flexible development of geoid surface \breps.
Extensibility of these frameworks
for the purposes of geophysical domain discretisation and model initialisation
is limited,
with for example GIS frameworks being built up from working on \twod raster fields.
Similarly, project files associated with GIS do not contain all of the information required to fully
constrain a spatial discretisation problem,
and moreover, it is not possible to include the high-level natural language functional descriptions proposed here.
As \cite{candygis} demonstrates though, GIS methods can benefit geophysical domain development, and their role is included in the schematic \cref{fig:process}.

Use of Spud enables a description of model option parameter space to be considered separately.
This is constructed in a schema file, a machine readable specification of which options are expected, their type and context, and how they should be read:
A formal grammar to be used to describe model constraints.
The
\cref{constraint:brep,constraint:hmetric,constraint:id,constraint:surfbounds,constraint:vmetric}
that fully describe the
geophysical domain discretisation problem
have been structured into a schema.
A schematic of the included components and their relationship to required constraints is shown in \cref{fig:block}.

This is a single hierarchical and formal description of the constraint space, and more generally the options available to the user in generating a mesh.
As illustrated in \cref{fig:process},
it is part of \shingle and
is central to how components of the approach interact with \brml files that describe a particular meshing problem.
At the simplest, highest level use of \shingle, this is transparent to the user.
For more advanced use and development, it provides a centralised and
language-based description of the constraint space that all other parts of
\shingle, and the geophysical mesh generation process, depend.

\subsection{Dynamic, hierarchical parameter description}
\label{sec:brep}
Just like the case of a numerical model,
there are a wide range of possible options in mesh generation,
even when restricted to geophysical problems.
The \brml schema builds on the general schema language for simulation models prepared in \cite{ham09},
to give an option-complete language for the mesh generation problem.
This is exactly the type of purpose Spud is intended for and other current models in development are adopting this approach to
formally describe model constraint spaces, like for example the new TerraFERMA model of \cite{wilson16}.

This caters for options which may be specified multiple times, at potentially varying levels of option hierarchy in multiple contexts.
For example, as the block diagram of \cref{fig:block} highlights,
a simulation domain can contain multiple geoid surfaces $\Gamma'$, each with potentially multiple boundary representation components (e.g. simple orientated polylines with identification).
\brml is an XML language, and by nature is hierarchical and extensible.
With this structure, and guided by what the schema permits (itself representing the
\cref{constraint:brep,constraint:hmetric,constraint:id,constraint:surfbounds,constraint:vmetric}),
it is easy to dynamically add, repeat, expand and remove options and groups of options whilst in context.

As an example, use of the Spud framework immediately provides access to the Diamond GUI
which enables easy editing and drafting of new domain discretisations.
This GUI uses the schema file (see \cref{fig:process}) to guide navigation of the option tree.
Through this the GUI knows to expect at least one definition of a geoid surface $\Gamma'$, for example, and a specification of a geoid metric $\mathcal{M}_h$
(and requires these from the user).
Additional geoid surfaces or more feature-rich boundary representation components are easily added and built up at a later stage, dynamically increasing the complexity of the mesh generation problem.

\subsection{Option tree cross-references}
Options are structured into a hierarchical tree within the \brml description.
The grouping of \constraints and decoupling (\cref{sec:decoupled})
are naturally structured in this way, as \cref{fig:block} highlights.
This is much like numerical simulation model options parameters,
which motivated the development of Spud and adoption of an underlying XML-based language.

In some cases there exist dependencies across the option tree,
and these are achieved through attribute names.
For instance, the choice has been made to centralise source dataset definitions.
These are named (e.g. `GEBCO2014' in \cref{fig:gui,fig:brml}) and this name referred back to whenever the data is required.
This is also used to assign potentially multiple \brep component sections to the
same named boundary identification (e.g. the `Coast' and `OpenOcean' named identifications of \cref{fig:gui,fig:brml}).

This also allows component \breps sections to be used multiple times.
This is required, for example, when distinct physical regions meet at an interface
(e.g. the open ocean meets an ice sheet) and share a boundary.
The component \brep section defining the interface can then be referred to out of the order defined by the hierarchy,
and from potentially separate parent geoid surface representation
$\Gamma'$
(where for instance $\Gamma'_o$ and $\Gamma'_i$ are setup to represent neighbouring geoid surface representations for the ocean and ice, respectively).

\subsection{Natural language descriptions}
Domains for geophysical simulations are typically described with reference to bounding lines on orthodromes such as meridians and parallels,
together with global or segments of contours such as a 0m coastline, for example.
More generally, geographic features are identified with a similar combination.
The Southern Ocean for example, is defined extending up from the Antarctic coast to the 60\degree S parallel, and the Atlantic and Indian Oceans divided at the 20\degree E meridian.

This is the natural way to identify bounds for geophysical models.
Setting up these geographic bounds and including all features contained within in a format suitable for meshing algorithms can be a time consuming, difficult to edit and repeat, ad hoc process.
\shingle automates this and from a basis of natural language definitions typically used in geophysical modelling studies.

The original consistent boundary representation generation approach described in \cite{candybrep} enabled sections of contours to be selected and domains extended meridionally to parallels.
This has been generalised significantly to allow a wide range of arbitrary bounds described with natural language definitions.
Moreover these can be defined multiple times, and in context with hierarchy available within the \brml description.
In the example presented in \cref{fig:gui,fig:brml}
the \brep can be seen to include two components:
a section of the Chilean coastline
and a second extending the domain out to a meridian at 7\degree W,
mirroring those in the description
\refdescriptionquotelink.

\subsection{Arbitrary and discrete descriptions}
More flexible functional descriptions can be made within the \brml written directly in Python.
This again in a relatively readable form, using primitives such as the positions `longitude' and `latitude',
or
Universal Transverse Mercator (UTM) coordinates `x' and `y'.
This can be used to describe an arbitrary orthodrome, for example.

In addition to this, the natural language basis can be supplemented with raw discrete data types such as orientated polylines from
the
GSHHG
database,
mapping databases
(e.g. the UK national \cite{os} resource)
or those developed directly in a GIS
as \cite{candygis} demonstrates,
bounding a domain to the complex UK coastline together with the fine man-made structures of Portland Harbour.
The high fidelity \brep is not only built up from components constructed on-the-fly
from
functional forms referencing geographic features,
but also
discretised forms containing an explicit description of
domain constraints, if needed (see \cref{fig:process}).
These are available through the central dataset section of the option hierarchy (\cref{fig:block}), and accessed from
local or distributed resources.

\subsection{Self-describing constraint options}
The constraint space description developed in the \brml schema is self-describing,
containing a verbose description of each option.
This information can presented alongside options in the GUI (see the top right of \cref{fig:gui}, for example) or reported for any option errors occurring at run time,
again from this centralised constraint space descriptor resource, the schema.
In this way the schema, and as a result the GUI, act as a manual, directly supporting users as mesh options are made.

From the developer's perspective,
this Spud based approach means new features can be added with minimal code changes.
The XML based structure means codes focus on patterns of options.
The schema defines what expected and the code loops through the hierarchy following well-defined patterns,
picking up options from a corresponding in-memory dictionary tree.

For the user, mesh generation with real fractal-like boundaries can be as simple as selecting a coastline segment by a bounding box and on the other side a bounding orthodrome, with choice of element edge-length metric (see \cref{fig:chile}).

\subsection{Provenance record}
A complete description of the domain discretisation problem is a fundamental requirement if an accurate record of
provenance is to be made, and this is provided by the \brml file.
These \brml files alone are themselves
easily parsable XML based problem description files, human-readable with structure.
This is focused on a textual natural language problem description and is lightweight as a result
such that changes are easily tracked with
version control systems such as Git and SVN.

Together with the problem description,
the \brml maintains
details of authors responsible for their creation, contact details,
comments including timestamped notes on past changes made in development (seen in \cref{fig:block,fig:gui}).
This is similar to the record kept within the global attribute metadata contained in NetCDF headers,
which is supplemented through operations performed on the data with tools such as
the Geospatial Data Abstraction Library \citep{gdal}.
The ADCIRC hydrodynamic circulation model \citep{westerink08} makes a record of this type of information in its NetCDF output,
inherited from its initialisation namelist files.
\shingle records this information in output where possible,
notably the high fidelity \brep,
supplementing it with a record of
the library release version and unique
repository abbreviated commit hash.
Unique identifiers of other libraries are also recorded,
such as the version of the meshing tool employed (e.g. Gmsh).

\section{Source data management}
\label{sec:data}
Data contributing to discrete domain characterisations
can be large in size, difficult to distribute efficiently and computationally costly to process.
The current version of the global bathymetry dataset \cite{gebco} containing only elevation is currently
$1.9\, \textrm{GB}$
in size, for example.
Efforts are growing to provide a complete provenance record of numerical model simulations,
with direct instructions from research funders requiring a research data management plan
\citep{nwodmp}
and in general, accountability from the public,
it is important to detail data source origin and content accurately.

Options for the management of mesh generation source data range from:
\begin{optionenum}
\setlength{\itemsep}{1pt}\setlength{\parskip}{0pt}\setlength{\parsep}{0pt}
\item Recast data into form suitable for distribution and share with \brml description.
\item Distribute processed datasets with \brml irrespective of size.
\item \label{dataoption3} Begin from a standardised raw dataset, and conduct potentially computationally demanding processing as needed.
\item \label{dataoption4} Refer to remote repositories of source data, such that data is downloaded and processed on demand.
\end{optionenum}

Often this data processing stage of the mesh generation process is not well-described, and difficult to reproduce,
with filtering, subsampling and agglomeration operations only loosely outlined.

Modern data descriptors support a record of provenance \citep[such as the `history' field embedded in NetCDF,][]{netcdf},
so it would be possible to record the filtering, subsampling and other processing here or within the \brml.

The purpose of the \brml description of constraints
is to provide an accurate description of the meshing problem.
It is not the intent to reinvent new standards for data description.
Along this line of design,
with a focus on provenance record and how data is handled,
and noting the computational demands and connectivity speeds that affect \cref{dataoption3,dataoption4} above will continue to improve in the future,
the approach is made to depend directly on raw, standard and potentially remote data sources.

\subsection{\opendap integration}
\label{sec:opendap}
The problem of efficient access to large remotely hosted data sources is tackled by
\cite{cornillon03} which describes \opendap (Open-source Project for a Network Data Access Protocol).
The protocol has since been adopted by many organisations
who host servers providing \opendap services.
This includes a large amount of environmental data in
the Global Change Master Directory (GCMD)
provided
over \opendap
by NASA\footnote{\url{http://gcmd.gsfc.nasa.gov}}.
Other data libraries such as the NOAA National Oceanographic Data Center%
\footnote{\url{http://data.nodc.noaa.gov/opendap}}
and
British Oceanographic Data Centre%
\footnote{\url{http://dods.bodc.ac.uk}}
are expanding the range of data delivered over \opendap.
Some \opendap servers additionally maintain a catalogue of other servers worldwide
such as the THREADDS host at Deltares%
\footnote{\url{http://opendap.deltares.nl}}.
Specific numerical simulation models too have their own dedicated servers to host output such as
the ocean models HYCOM%
\footnote{\url{http://tds.hycom.org}}
and
ROMS%
\footnote{\scalebox{0.97}[1.0]{\url{http://megara.tamu.edu:8080},~\url{http://tds.marine.rutgers.edu}}}.

This has typically been applied to sharing geophysical model output data in combination with
the \citep[NetCDF][]{netcdf} and Climate and Forecast \citep[CF,][]{gregory03} metadata standards \citep{hankin10},
for intercomparisons and post-processing analysis.
Here we apply \opendap to model initialisation.
In \shingle, this \opendap negotiation is achieved using the standard Python library pydap.
In this way \shingle can request fundamental operations
are applied to distributed datasets before they are delivered for further processing,
picking out required fields and regions of interest to reduce the size of data communicated.
A description of further processing such as subsampling and filtering is then maintained in the \brml and
executed through standardised Python wrappers to established geospatial tools such as \cite{gdal}.
A reference in place for the \cite{gebco} data source hosted on the NASA/JPL ECCO \opendap server
is made in \cref{fig:brml},
where the region of interest
(for cropping on the remote server)
is automatically established by its use further down in the tree.

Keeping the \brml focused on problem description, with references to source data,
ensures it is lightweight and portable.
Iterative adjustments to the mesh generation are also
then made with changes to descriptions rather than data.
Furthermore, these are then easily managed in version control systems.

This additionally ensures the verification test engine is lightweight
and apart from a dependence on standard software libraries,
and a connection to \opendap servers,
is self-sufficient
and can be easily be setup and used independently.

Constraints built from distributed resources are encouraged,
but to engage with existing mesh generation workflows and as a pragmatic solution,
source files can be cached or local files used directly (see \cref{fig:process}).

\subsection{Self-consistent \brep development}
\shingle applies the self-consistent approach to mesh generation developed in \cite{candybrep}.
Within the \brml description this is emphasized
through
a central
data source definition
(seen in \cref{fig:block,fig:gui,fig:brml}),
rather than external sources brought in directly
at different levels in the hierarchy
and correspondingly in the
generation process (\cref{fig:block}).
It is then easier to ensure datasets and their component fields undergo the same pre-processing
to generate high fidelity constraints that are consistent, and a solution spatial discretisation that is self-consistent.

Data used to construct the spatial domain discretisation
is commonly a DEM describing a surface through perturbations from a reference geoid surface
(e.g. to establish a geoid surface \brep),
but is not limited to this form, with for example \cite{candybrep}
developing a mesh optimised to the mean track of the ACC, based on currents in the Southern Ocean.

\section{\libshingle, the Shingle library framework}
\label{sec:framework}
\begin{figure}[!h]
\footnotesize{(a)}\vspace{-1.6ex}
\begin{pythoncode}
from shingle import SpatialDiscretisation, Dataset, Boundary
# Set up constraints
R = SpatialDiscretisation(name='NorthSea')
R.SetProjection('UTM', -3, 52) # alternatively zone='30U'
gebco = Dataset(type='raster', source='opendap', url='...', region=[-12,14,45,62])
coast = Boundary('coast', id=3)
S = R.AddSurface()
S.AddBoundaryComponent(source=gebco, contour='ocean0m', id=coast)
   ...
M = R.Discretisation()
M.Save('NorthSea.msh')
\end{pythoncode}
\vspace{-0.3ex}\footnotesize{(b)}\vspace{-1.6ex}
\begin{pythoncode}
# Modify boundary representation output projection
import pyproj
p = pyproj.Proj('+proj=utm +zone=30U +ellps=WGS84 +datum=WGS84 +units=m")
R.SetProjection(p)
R.Save('NorthSea_UTM30U.brml')
\end{pythoncode}
\vspace{-0.3ex}\footnotesize{(c)}\vspace{-1.6ex}
\begin{pythoncode}
# Simple parameter sweep example
from shingle import Load
R = Load('Weddell_Sea.brml')
S = R.GetSurface('SouthernOcean')
B = S.GetBoundaryComponent('OpenParallel')
for latitude in [float(x) for x in xrange(-75,-65,2)]:
  B.ExtendToParallel(latitude)
B.Save('Weddell_Sea_and_
\end{pythoncode}
\vspace{-1.8ex}
\caption{
Example interactions with the Shingle Python library \libshingle.
(a) Using natural language constructs native to Shingle, counterparts to \brml entries.
(b) Together with objects native to external libraries.
(c) Loading, extending and saving descriptions from \brml.
}
\label{fig:python}
\end{figure}
\subsection{Built on standard libraries}
\label{sec:standard}
The library \libshingle is written in Python and uses standard libraries for operations where possible.
It can simply be used transparently through the \shingle executable to interpret the constraints specified in \brml file descriptions.
For lower level more advanced use building up constraints for more complex setups or
in prototyping natural language objects for automating the inclusion of new geographic features,
interaction can be made directly with the \libshingle library as \cref{fig:process} illustrates.

Mirroring the \brml constraint description (overviewed in \cref{fig:block}),
the library contains natural language based objects that can be built up
in code to construct components of a mesh generation problem,
including boundary representations and element edge length metrics.
The mesh problem can then be solved under these constructed constraints all within a Python context.

\libshingle uses the open source Python shapely library
(refer to \cref{fig:process})
to handle polyline imports and manipulations.
The Scientific.IO library is relied on to efficiently process raster NetCDF files.
The homeomorphic projections to the charts required in the mesh generation process \cite[see][]{candybrep},
such as $\xi$ of \cref{brep} are interpreted and managed by the Proj.4 Python library pyproj.
Geospatial operations can be made by both high-level \shingle objects, or built up with
GDAL
operations through its Python osgeo interface.

Although the use of external libraries may require updates to \shingle in the future to maintain compatibility,
this is minimal compared to the benefits
of using standardised implementations (\cref{tenet:standard}), that have community effort to ensure ongoing support with operating systems and interaction with other software and methods.

\subsection{Low-level interaction through Python objects}
In addition to the ongoing support from standard libraries in high-level use,
\shingle has been written to interact directly with external libraries.
Objects such as pyproj projections, GDAL operations, surface and polyline descriptions can be used interchangeably with \libshingle.
An example bringing in a UTM projection setup externally using the standard library pyproj is shown in \cref{fig:python}(b).
This supplements the high-level text-based natural language definitions available in the \brml,
and a route to adding new high-level
boundary representation \brml objects to \libshingle as needed.

\subsection{Efficient parameter space exploration}
In developing a new application study applying a numerical simulation model,
it is common to iterate on a spatial discretisation until it is optimum and fit for purpose.
This involves small changes in the constraints, exploring parameter space often through
a loose bisecting binary search algorithm.
This process can be rigorously implemented and automated with \libshingle,
where modifications are guided by the schema describing the formal grammar of the constraint space through libspud.
\Cref{fig:python}(c) illustrates a simple template to modifying and generating a range of \brml mesh descriptions.
The solution mesh discretised domains can be generated in the same way,
and this could further be used to initiate numerical simulation runs.

This algorithmic formulation of constraints is easily extended to enable
complex operations that are difficult to achieve with other approaches.
For example, the loop of \cref{fig:python}(c) is trivially extended to include a search algorithm exploring a parameter space to converge a domain discretisation on a required total number of nodes and hence degrees of freedom.

Being an XML based language, the \brml descriptions can also be simply interrogated and modified directly
with standard XML libraries.  This interaction is highlighted separately in \cref{fig:process}.

\begin{figure*}[!h]
\begin{center}
\includegraphics[width=\textwidth]{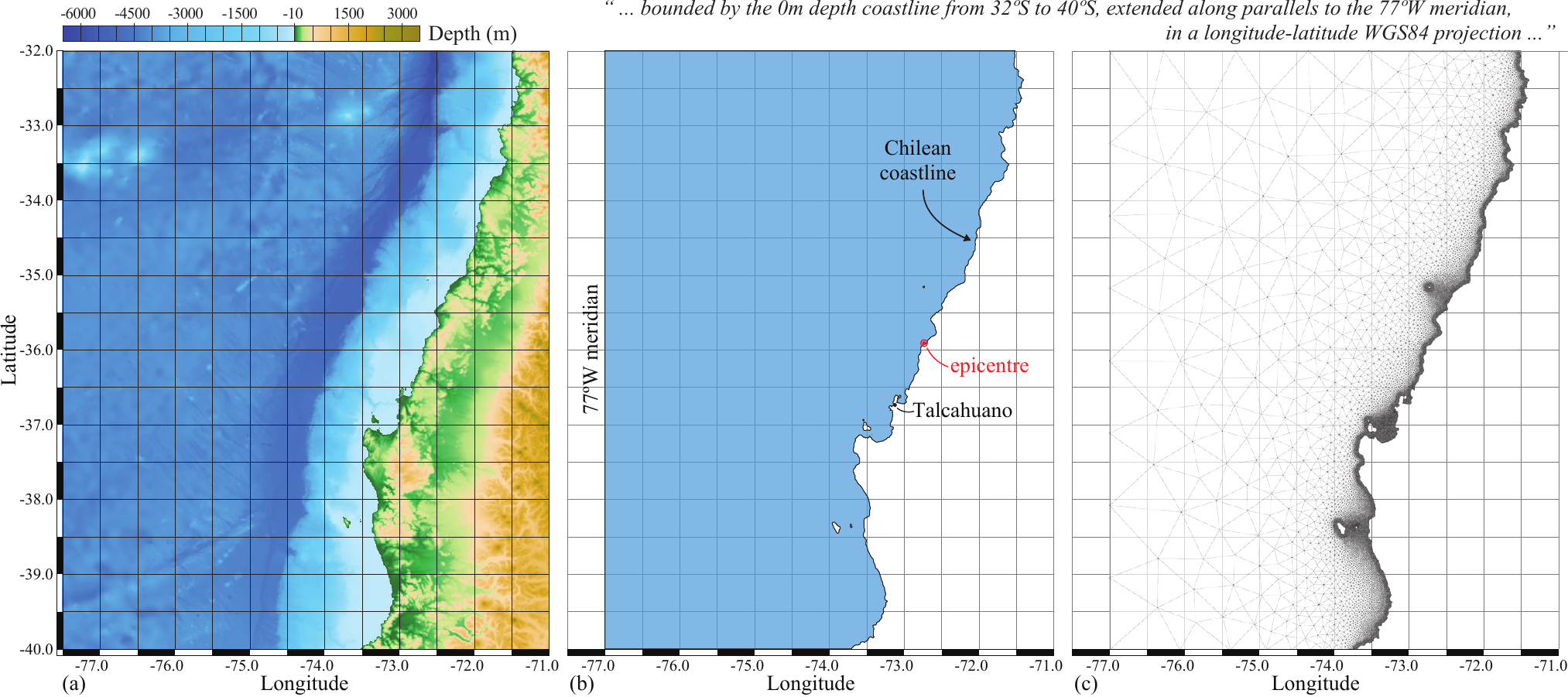}
\end{center}
\vspace{-1.8ex}
\caption{
Example simulation domain for modelling ocean wave propagation and tsunami inundation
in the 2010 Chile M8.8 earthquake, centred at
35.9\degree S 72.7\degree W,
approximately 100km north of Talcahuano.
This domain is relatively simply described by
\refdescriptionquotelink
in \cref{sec:introduction}
with constraints formally defined
by the \brml of \cref{fig:brml}
(with some further description and corresponding formal \brml to constrain spatial resolution).
Generation is a simple matter of translating the former into the latter under the formal grammar, with both being human-readable descriptions.
\shingle automatically handles the details of defining a high fidelity \brep
$\Gamma'$ in (b)
from the \cite{gebco} DEM (a)
and, notably here, includes island features
to give a geoid surface representation with non-zero genus \citep[following the approach of][]{candybrep}, and further to automatically produce a simulation-ready meshed spatial discretisation
$\mathcal{T}_h$ in (c).
}
\label{fig:chile}
\end{figure*}

\section{Model, method and data interaction and interoperability}
\label{sec:interaction}
\shingle has been built with modules for high-level interactions,
with established tools used in mesh generation.
These are highlighted in \cref{fig:process}, with a core link to the Gmsh library of meshing algorithms.
Where possible interaction is achieved through standardised Python APIs,
such as the \cite{trianglepython} for the Triangle \citep{triangle} library of Delaunay mesh algorithms.
High fidelity boundary representation can be output in Gmsh format
using a specific format writer developed within a collection of writer modules prepared within \shingle.

Similarly, fields supporting a meshed domain
(e.g. initial full-field temperature state)
can be output as unstructured VTK files, using a format writer extending standard VTK libraries.
Data is written and stored efficiently in an XML based data format containing blocks of binary data compressed using the zlib library.

\subsection{Model format writers}
Models with non-standard data formats are supported through specific format writers.
This modular approach enables new format writers (and readers) to be added as needed.
As examples, \shingle includes modules to prepare initialisation files for
the ADCIRC hydrodynamic circulation model and H2Ocean shallow water equation model.

As well as writing mesh solutions, the output writers are used for validation purposes and in the general purpose efficient prototyping (\cref{tenet:efficient}).
Output can be prepared for viewing alongside source data in geospatially valid context provided by GIS frameworks, with for example the resulting mesh and discrete bounds overlaid over DEMs directly within GIS \citep[see][]{candygis}.
This is useful for a visual evaluation of conformity, to see how well geographic features are represented.
For large discretisations, visualisations tools designed specifically for efficiently handling large unstructured datasets can be employed, such as
Paraview\footnote{\url{http://www.paraview.org}},
which is directly supported by \shingle using VTK.

Interaction at different levels is important to ensure a hierarchy of automation \cref{tenet:automated}.
Particularly challenging meshing problems can, for example, easily be offloaded to more capable dedicated resources.

For quick visual inspection purposes, \shingle can automatically output an image of the geoid surface mesh discretisation.

\subsection{Input readers}
Parallel to the writer modules, \shingle includes readers.
These are used to interact with meshing libraries where needed, loading in output mesh discretisations produced by Gmsh on-the-fly, for example.
Additionally this can be used to support a wider range of data sources and initialisation.
Standard data in NetCDF and shapefile form can be read.
Readers here can import more complex heterogeneous data, including GIS projects with multiple layers containing a wide range of data types, for example.

\subsection{Embedding in model codes}
As a Python library unifying \brep constraint and solution, \libshingle makes it possible to incorporate complex domain discretisation of real geophysical domains in overarching model control scripts, which is where development of new cutting-edge models is headed
\citep[see for example,][]{firedrake,omuse}.
In this way the model supplements the problem constraints sent to \libshingle (see \cref{fig:unified}),
dependent on numerical discretisations employed
in the simulation model, and
the \brml would be truly independent of specific models, a pure description of the boundary representation, resolution and identification.
Moreover, interaction through the library enables models to
handle the output discretisation directly as the Python objects constructed by \shingle, rather than an intermediate file object.

As \cite{omuse} demonstrates, complex multi-model Earth system models can be created and coupled, and interactively monitored,
on potentially a heterogeneous array of computational resources,
all coordinated from a central a Python interface.
\libshingle brings domain discretisation in real geometries to these type of extensible Earth system modelling frameworks.

\section{Verification and discretisation validation}
\label{sec:verification}
\begin{figure*}[!h]
\begin{center}
\includegraphics[width=\textwidth]{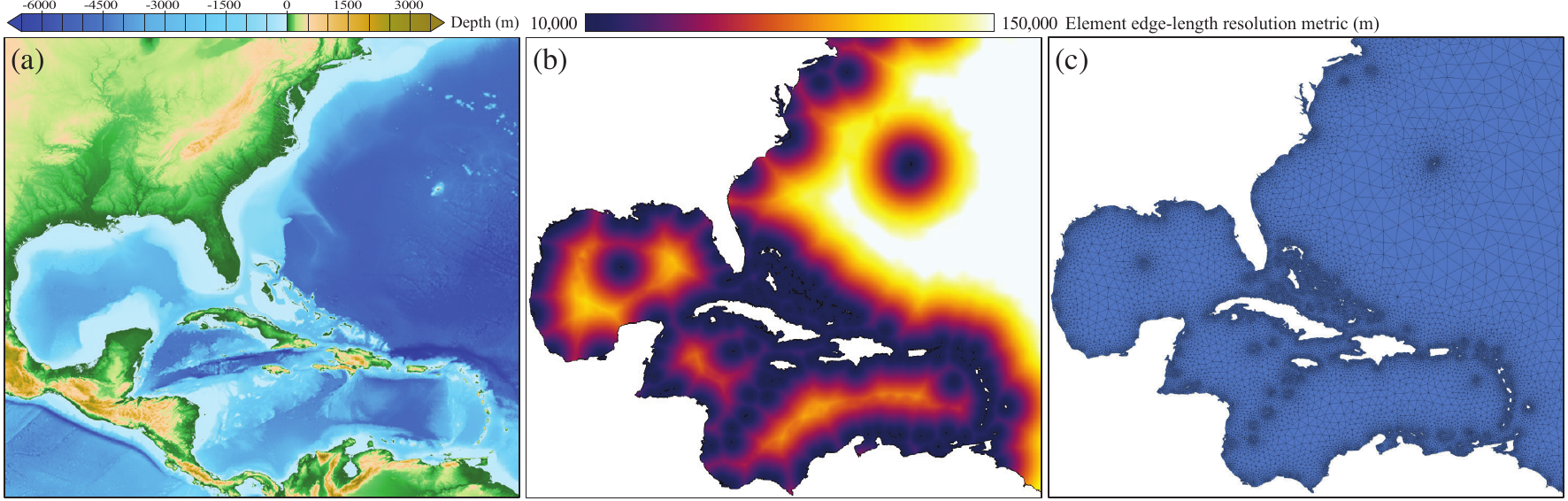}
\end{center}
\vspace{-1.8ex}
\caption{
Simulation domain focused on the Caribbean Sea basin.
(a) \cite{gebco} DEM.
(b) Surface geoid element edge-length resolution metric
$\mathcal{M}_h$ developed as a function of (a).
(c) Surface geoid \brep
$\Gamma'$
in blue,
overlaid with multi-scale spatial discretisation
$\mathcal{T}_h$.
}
\label{fig:caribbean}
\end{figure*}

\begin{figure*}[!h]
\begin{center}
\includegraphics[width=\textwidth]{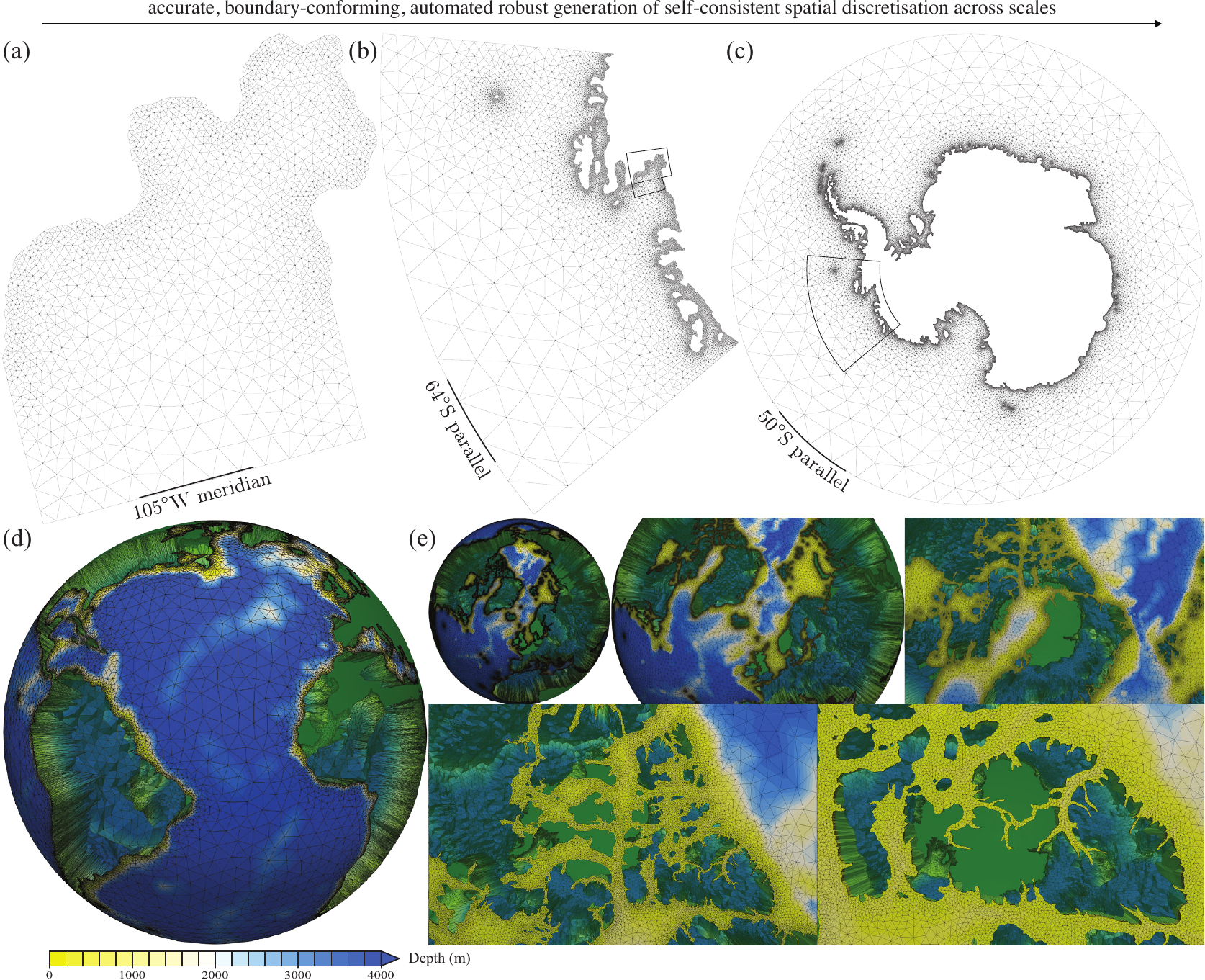}
\end{center}
\vspace{-1.8ex}
\caption{
A selection of further example geophysical domain discretisations
straight-forwardly
described in \brml and
automatically
constructed using \shingle.
(a) $\mathcal{T}_h$ of the Pine Island Glacier ice shelf ocean cavity from ice-bedrock grounding line extended out to the
105\degree W meridian.
(b) The Amundsen Sea region in West Antarctica extended out to the 64\degree S parallel.
(c) The Southern Ocean Antarctic continent landmasses, from ice grounding line to 50\degree S parallel,
built from a high fidelity \brep containing 348 automatically identified islands.
(d) The full $\mathcal{T}$ of the global oceans,
with a radial scaling of 300 to exaggerate the vertical extent of the discretised shell
and land regions shaded green.
(e) Zoomed in regions focusing on the complex Canadian Arctic Archipelago west of Greenland around Ellesmere and Baffin island.
(a)--(c) are generated from the \cite{gebco} DEM and presented under a
orthographic projection centred on 90\degree S,
and (d)--(e) from RTopo \citep{rtopo} and viewed in a Cartesian frame.
These contain a multi-scale of spatial resolutions, with element edge-lengths parallel to the geoid in these examples,
specified through $\mathcal{M}_h$,
ranging from
$2\textrm{km}$
to
$500\textrm{km}$.
Vertical layers in (d), specified through
$\mathcal{M}_v$,
vary from
$2\textrm{m}$
to
$500\textrm{m}$,
under differing regimes in a generalised hybrid coordinate system described further in \cite{candybrep},
and leads to a mesh containing 8,778,728 elements and 35,114,912 spatial degrees of freedom under its discontinuous Galerkin finite element discretisation.
Along with other examples presented in \cref{fig:challenge}(c)-(g),
these are part of the test suite accompanying the library.
}
\label{fig:discretisations}
\end{figure*}

A suite of verification tests are provided together with \shingle, along with the automated test engine detailed in
\cref{sec:continuousverification}.
A selection of geophysical domain discretisations
described in \brml that form part of the test examples
are shown in \cref{fig:chile,fig:caribbean,fig:discretisations,fig:challenge}.
Each test is evaluated using validation tests built into \shingle and their \brml descriptions, as outlined in
\cref{sec:self-validation}.
The test engine can be used to verify a new install, and flexibly to support iterative mesh drafting and prototyping (\cref{tenet:efficient}).
\subsection{Self-validation}
\label{sec:self-validation}
\begin{figure*}[!h]
\begin{center}
\includegraphics[width=\textwidth]{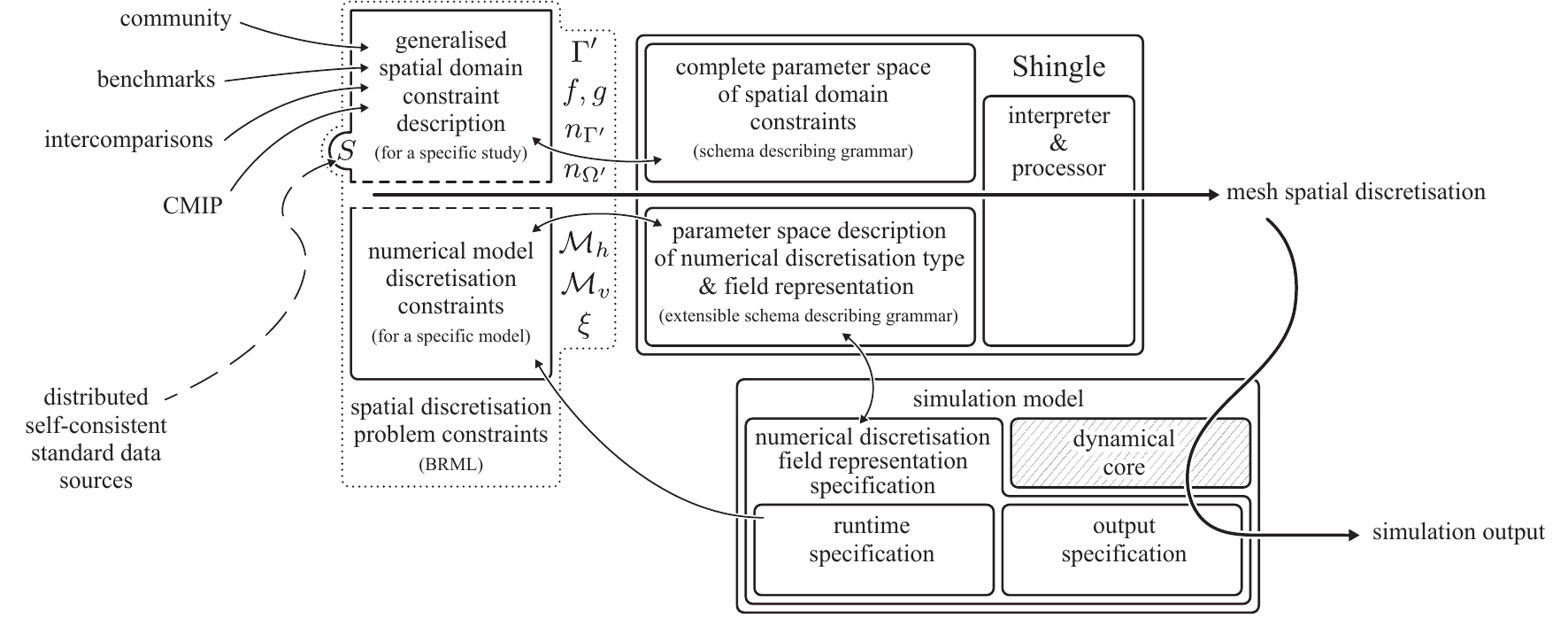}
\end{center}
\vspace{-1.8ex}
\caption{
Framework for generalised spatial domain discretisation for geophysical model simulations.
A formal spatial domain constraint description
(a model-independent grouping of high-level directives describing key geospatial boundaries and features, required spatial resolution and source datasets)
for a specific study (e.g. the geography to include in a CMIP intercomparison study) is
joined with specific constraints from a simulation model, depending on its internal numerical discretisations and field representations
(e.g. following Gridspec \citep{gridspec}, or a UFL description \citep{ufl}).
These constraints are used by the interpreter \shingle to produce, in a robust, automated, repeatable process, a model-specific mesh spatial discretisation.
Moreover, the latter description is further used to specify numerical simulation output representation
(as CMIP uses Gridspec).
}
\label{fig:unified}
\end{figure*}
Validation of the mesh generation process is achieved in four ways.
Firstly,
with reference to the formal grammar of the constraint space,
a degree of self-validation can take place on-the-fly as mesh options are built up.
Following rules described in the schema,
only some options are available
and certain combinations permitted.
Unlike with namelist descriptions,
or ad hoc collections of data,
the user does not need to wait until running \shingle before receiving feedback on option validity.
Available options are limited dynamically following the constraints and option selections.
Moreover, with information from the schema on the mesh generation problem, it is possible to identify which options are required for the problem to be complete.
The creation of a new \brml file immediately requires a name, type and options to be completed for at least one geoid surface representation and a geoid metric.
The GUI highlights which required options remain to be completed (see \cref{fig:gui}).
This is particularly useful to users new to mesh generation.

Secondly, the required `type' option classifies the mesh and checks at runtime it is suitable for the intended simulation.
A `shallow water' model requires only a surface geoid discretisation
$\mathcal{T}_h$ for example, whilst a full \threed mesh is needed in other simulation types.
This is a sanity check to ensure the mesh generation problem is fully constrained for the intended purpose, beyond the fundamental \constraints.

Thirdly, a parsing stage following application of a meshing algorithm eliminates commonly found issues in output mesh descriptions,
ensuring structural integrity.
For example, additional lone, unconnected boundary elements are removed in this step to ensure the discretised output mesh is as expected.
Meshing algorithms do not usually possess information on underlying numerical discretisations, and it is also possible elements are generated that `tied' to boundary conditions, with no independent free unknowns.
This type of problem in the spatial discretisation is often difficult to identify, only being picked up at runtime, or through careful visual inspection.
This parsing is an opportunity to identify and process these at this stage.
Numerical simulation codes are sometimes accompanied with standalone mesh checking tools to support initialisation stages
(e.g. the MechChecker.F90 utility for the ADCIRC model),
and visual interfaces can be used for manual inspection and editing,
such as the Show Me tool provided alongside Triangle \citep{triangle} and the GUI of Gmsh \citep{gmsh}.
This is part of the mesh generation process and,
if possible, better handled automatically following \cref{tenet:automated}, as proposed.

Lastly, the fourth approach to validation is through explicitly defined
expected boundary representation and discretised mesh characteristics.
Like the initial consistent approach of \cite{candybrep}, the intermediate high fidelity
boundary representation is compared at a raw level.
Being a deterministic process, deviations are only expected as a result of
depending on \shingle library version and behaviour, source data and potential \opendap response, machine precision and the originating \brml description.
At this stage of the meshing process, this has been supplemented with a test on the area within the bounds of the high fidelity geoid \brep $\Gamma'$.

On the discretised output, the tests include
simple
lower and upper
bounds on output geoid mesh node and element numbers,
the number of boundary elements,
and element circumspheres to check
adherence to metric constraints.
The degree of representation is examined comparing the high fidelity geoid \brep surface area to its corresponding discretised form.
Boundary complexity is measured through the overall Minkowski fractal dimension.

This provides a means for users to easily specify what should be expected in the discretised output,
to ensure the accuracy required in \cref{tenet:brep,tenet:metric,tenet:region}.
Testing built in to the mesh generation process, further automates the process.
It is also important to ensure \cref{tenet:provenance}: provenance, that the solution mesh is the same (within prescribed tolerances) as that that has been generated in the past, and potentially by others on different systems.

A self-validating description provides \cref{tenet:standard}: a standardisation of interaction with the descriptions themselves.
Users can immediately begin building on and improving the work shared by others, having been able to check the descriptions give a solution expected by the creator.
This eliminates ad hoc or purely qualitative measures of conformity and reinforces the provenance record of the mesh generation process.

This is important when these then form key components of critical studies, such as the
coupled climate and Earth system models run for
internationally
coordinated
model intercomparisons,
such as
CMIP and CORE
\citep{meehl07,taylor12,coreii},
that form the basis of
reports compiled by the IPCC.

Models containing unstructured meshes with conforming boundaries are now starting to be used in such large-scale international research efforts
\citep[e.g. FESOM,][]{sidorenko14}.
This approach provides the
full provenance, reproducibility and complete constraining descriptions of the significantly more complex spatial discretisations supported by these models.

\subsection{Continuous verification}
\label{sec:continuousverification}
To ensure \shingle as a whole continues to behave as expected for all users and on all systems, it contains a verification test engine.
This processes a suite of key meshing problems, which are then automatically evaluated following the validation tests defined in their \brml description.
Since the \brml descriptions are self-validating, the addition of new tests to the suite is simply a matter of adding the problem description file to a test folder of the source code.
Testing is often a secondary consideration to new feature implementation, so it is important the extension of testing suite is as simple as possible.

This can simply be run at the time of a new installation,
following the upgrade of required libraries or the operating system,
or routinely as part of a commit-hook buildbot with dedicated resources to continuously verify new code pushed to a \shingle development code repository \citep[see, for example,][]{farrell11}.
Being built on standard libraries, it could further form part of an automated
wider system framework validation, for the above climate intercomparison projects, for example, reproducing the entire process from initialisation to post-processing, on demand.
Alternatively, the engine can be used to drive an efficient drafting and prototyping workflow (\cref{tenet:efficient}) with updates to mesh generation problems automatically processed and tested,
to support an iterative domain discretisation process.

\conclusions
\label{sec:conclusion}
This research has developed a high-level abstraction to mesh generation for domains containing complex, fractal-like coastlines that characterise those in numerical simulations of geophysical dynamics,
together with a
compact, shareable and necessarily complete description of the domain discretisation.

The approach is designed to be accessible to a wide range of users and applications.
This begins at a simple standalone GUI-driven one way workflow,
where users are guided through the option parameters required to constrain the domain discretisation problem.
Options are presented in context through the hierarchical tree structure with documentation automatically provided alongside.
Moreover, the use of a human readable XML format and introduction of high-level natural language based geographical objects give \brml problem constraint descriptions that closely follow those presented in literature and shared by scientists.
The example
built up from the description
\refdescriptionquotelink,
to \brml in \cref{fig:brml,fig:gui},
followed by the construction of
the high fidelity \brep and resulting spatial discretisation
shown in \cref{fig:chile}, highlights how the problem of generating a domain bounded by a complex coastline defined by a depth contour and three orthodromes,
common in tsunami modelling studies, is trivially constructed and solved using \shingle.

This is easily built on and extended to larger and more complex problems.
High-level objects automate processing of multiple, potentially complex geospatial features.
\brml descriptions are easily shared and XML sections cut and pasted to combine descriptions and build up complexity.
New high-level objects and processing can be prototyped directly in Python to later join the core \libshingle operations library.
Corresponding natural language based objects are available through the Python API, meaning domain discretisation can be achieved directly
and purely in native Python code, for complex setups, direct integration with numerical simulation codes, or interactive sessions or
Jupyter notebooks%
\footnote{\url{http://jupyter.org}}.
Both the \brml file descriptor and modular \libshingle are extensible.

Extending the tsunami example shown in \cref{fig:chile}, this robust and automated approach could form part of a real time warning system using unstructured spatial discretisation,
with a domain
created on-the-fly, centred around the earthquake epicentre, in a direct response to measurement by GPS seismic monitors.

Recognising the domain discretisation process is becoming more challenging
and more difficult to document completely, such that others can reproduce, has been central
to steering this approach.
Progress is focused on the nine tenets of geophysical mesh generation summarised in
\cref{fig:tenets}.
One result of this is that \shingle treats the mesh generation as a model problem.
Strategies from
numerical simulation model development have been adopted and modified
to formalise the description of the heterogeneous geophysical mesh generation constraints,
such that they provide an accurate and complete description
(\cref{tenet:brep,tenet:metric,tenet:region})
in a standardised language-based XML form
(\cref{tenet:standard}).
This compact text-based description easily affords
a record of changes in the development of a domain discretisation
(\cref{tenet:provenance})
and through the \brml grammar ensures it is always a complete description and therefore reproducible.
The model-based approach manages the range types of parameters
(which have diversified with the use of flexible unstructured discretisations)
and supports users in their preparation,
to allow for efficient drafting and prototyping
(\cref{tenet:efficient}).
With options managed in a structured hierarchical tree,
complex discretisations can be built up logically
(\cref{tenet:automated})
and scaled up
(\cref{tenet:scales}).

The creation of the \brml file \brep description is not intended to reinvent standards.
It is not a new data descriptor,
for orientated vector paths or \twod raster data, for example.
There exist standards already that tackle these challenges well.
It is rather a new problem descriptor,
like those for Fluidity \citep{piggott08} and the TerraFERMA model of \cite{wilson16},
for fully describing the mesh generation problem specifically for geophysical model domains,
following the approach that this requires solving the same types of challenges involved in numerical model setup,
that makes significant progress in meeting the
tenets of \cref{fig:tenets}.

The consistent approach of \cite{candybrep} is adopted,
with an emphasis on producing a self-consistent high fidelity description and resulting output domain discretisation.
Consistency is additionally encouraged through a centralised definition of the source data and processing in the \brml description
(see \cref{fig:block,fig:gui,fig:brml}).
Use of decentralised, distributed datasets,
efficiently accessed using \opendap,
ensures the discretisation uses exactly the same source data
on every processing instance.

Verification and discretisation validation is achieved at multiple points throughout the process.
The formal grammar of the \brml,
imposed
by the schema,
enforces valid inputs and
provides initial option checking.
This framework and interaction with the schema using the libspud library additionally enables new
self-validating user interfaces to be written.
With expected mesh validation measures included in the \brml descriptions,
discretisations are automatically validated
and
continuous verification of the library is easily obtained.

With the dependable, robustly verified library \libshingle
for high-level abstractions for geophysical mesh generation,
it is easily applied to develop interactions with other frameworks and models,
such as GIS, as described in \cite{candygis}.
Critically, with the standalone \libshingle library,
these are easier to maintain and better insulated to API changes in other codes.

It does not immediately solve the mesh generation constraint problem in general, since numerical simulation models use a wide range of mesh types and numerical discretisations.
It has however, been designed with this in mind,
with low-level structures that are extensible,
to accommodate additional mesh types for example,
and high-level constructs that are applicable to all geophysical models.
Arguably the \emph{`holy grail'} of domain initialisation for geophysical models,
characterised by the
\constraints
following the development
\cref{fig:block},
is a grouping of high-level directives describing bounds (including key geospatial features to capture), required spatial resolution and source datasets
that can be interpreted by any model,
each dealing with the discretisation depending on the field representations within the model (\cref{fig:unified}).
\shingle provides an extensible platform to achieve this, focusing on general, natural language based, model-independent descriptions of domain descriptions, that can be shared and used for different models.
\libshingle additionally provides a means to interpret these descriptions such that this part of the process can be included in numerical simulation code, with the \brml constraints supplemented by those imposed by the simulation model, such as specific numerical discretisation choice (e.g. to use hexagonal over triangular prism elements), or
ensuring a minimum degree of representation in maintained between bounds (e.g. within narrow river channel networks).

\section*{Code availability, distribution and licensing}
\label{sec:distribution}
The \shingle computational research software library, developed as part of this study, is
available at
\url{https://github.com/shingleproject/Shingle},
with further information at
\url{https://www.shingleproject.org}.
This is accompanied by
a manual,
a suite of example domain discretisation \brml descriptions
and the verification test engine presented in \cref{sec:verification}.

All components of the \shingle package
which have been under continued development since 2011
are free software, being released under the GNU General Public License version 3.0.
Full details of the license, including the compatible copyright notices of third party routines included in the package, are included in COPYING in the source distribution.

\section*{Acknowledgements}
The authors wish to acknowledge support from the Netherlands Organisation for Scientific Research (NWO, grant number 858.14.061),
and also thank
{Gerben de Boer} for discussions on \opendap and its adoption within the Netherlands and more generally by the wider scientific community.

\end{document}